\documentclass[10pt, conference, compsocconf]{IEEEtran}   	
\usepackage{graphicx}				

\usepackage{color,flushend}

\usepackage{amssymb, amsmath}
\usepackage{mathtools}
\usepackage{booktabs}
\usepackage{verbatim}

\usepackage{clrs+} 
\usepackage{algorithm}
\usepackage[noend]{algpseudocode}
\usepackage{varwidth}
\usepackage{multirow}
\usepackage{rotating, xcolor}
\usepackage{subfig}
\usepackage{enumitem}
\usepackage[hyphens]{url}
\usepackage{float}
\newfloat{algorithm}{t}{lop}

\usepackage{tcolorbox}
\definecolor{mycolor}{rgb}{0.122, 0.435, 0.698}


\algnewcommand{\LineComment}[1]{\State \(\triangleright\) #1}

\newcommand{\erdosrenyi}{Erd\H os-R\'{e}nyi }

\newcommand{\qc}{\c{c}}

\usepackage{mathtools}

\DeclarePairedDelimiter\floor{\lfloor}{\rfloor}

\newcommand{\mA}{\mathbf{A}}

\newcommand{\dnzc}{\id{nzc}}

\newcommand{\dnnz}{\id{nnz}}

\newcommand{\dth}{th}

\title{A work-efficient 
parallel sparse matrix-sparse vector multiplication algorithm}

\author{Ariful~Azad,
Ayd\i n Bulu\qc~\IEEEmembership{}\\
  
  \{azad,abuluc\}@lbl.gov\\ 
  Computational Research Division \\
  Lawrence Berkeley National Laboratory \\
}

\date{}	

\begin{document}
\maketitle

\begin{abstract}
We design and develop a work-efficient multithreaded algorithm for sparse matrix-sparse vector multiplication (SpMSpV) where the matrix, the input vector, and the output vector are all sparse. SpMSpV is an important primitive in the emerging GraphBLAS standard and is the workhorse of many graph algorithms including breadth-first search, bipartite graph matching, and maximal independent set. As thread counts increase, existing multithreaded SpMSpV algorithms can spend more time accessing the sparse matrix data structure than doing arithmetic. Our shared-memory parallel SpMSpV algorithm is work efficient in the sense its total work is proportional to the number of arithmetic operations required. The key insight is to avoid each thread individually scan the list of matrix columns. 

Our algorithm is simple to implement and operates on existing column-based sparse matrix formats. It performs well on diverse matrices and vectors with heterogeneous sparsity patterns. 
A high-performance implementation of the algorithm attains up to 15x speedup on a 24-core Intel Ivy Bridge processor and up to 49x speedup on a 64-core Intel KNL manycore processor.
In contrast to implementations of existing algorithms, the performance of our algorithm is sustained on a variety of different input types include matrices representing scale-free and high-diameter graphs. 
 

\end{abstract}
\section{Introduction}

Sparse matrix-sparse vector multiplication (SpMSpV) is an important computational primitive with many applications in graph algorithms and machine learning.
The SpMSpV operation can be formalized as $\mathbf{y} \gets \mA \mathbf{x}$ where a sparse matrix $\mA$ is multiplied by a sparse
vector $\mathbf{x}$ to produce a (potentially sparse) vector $\mathbf{y}$. 
Due to lack of applications in traditional scientific computing, the research community has not paid much attention to computing SpMSpV efficiently.
It is possible to interpret SpMSpV as a special case of sparse matrix-matrix multiplication where the second matrix has dimensions $n\times 1$.
While this formulation can be relatively efficient for computing SpMSpV sequentially, for example by using Gustavson's SpGEMM 
algorithms~\cite{gustavson1978two}, it is not a good fit for computing SpMSpV in parallel. This is because there is often little work in each SpMSpV 
operation, necessitating novel approaches in order to scale to increasing thread counts.

The computational pattern in many graph algorithms involves transforming a set of 
active vertices (often called ``the current frontier'') to a new set of active vertices (often called the ``next frontier'').
Such graph algorithms, which are often called ``data-driven' algorithms''~\cite{lenharth2016parallel}, are harder to parallelize because the work changes dynamically
as the algorithm proceeds and there is often very little work per transformation.  
This ``frontier expansion'' pattern is neatly captured by the SpMSpV primitive: the current frontier is represented with the input 
vector $\mathbf{x}$, the graph is represented by the matrix $\mA$ and the next frontier is represented by $\mathbf{y}$.
For this reason, SpMSpV is the workhorse of many graph algorithms that are implemented using matrix primitives,
such as breadth-first search~\cite{bfs:11}, maximal independent sets~\cite{kdtsejits2013}, connected 
components~\cite{ekanadham2016graph}, and bipartite graph matching~\cite{azad2016distributed}. This
makes SpMSpV one of the most important primitives in the upcoming GraphBLAS~\cite{mathgraphblas16} standard (\url{http://graphblas.org}). 

Even seemingly more regular graph algorithms, such as PageRank, are better implemented in a data-driven way using the SpMSpV
primitive as opposed to using sparse matrix-dense vector multiplication. 
This is because SpMSpV allows marking vertices ``inactive" using the sparsity of the input vector, as soon as its value converges (i.e. stops changing).
Finally, local graph clustering methods such as those based on
the Spielman-Teng algorithm~\cite{spielman2013local} or the more practical Andersen-Chung-Lang (ACL) algorithm~\cite{andersen2006local} 
essentially perform one SpMSpV at each step.

In the area of supervised learning, SpMSpV becomes the workhorse of many support-vector machine (SVM) implementations that use the sequential minimal 
optimization (SMO) approach~\cite{chang2011libsvm}. In this formulation, the working set is represented by the sparse matrix $\mA$ and the sample data is 
represented by the sparse input vector $\mathbf{x}$.
SpMSpV is also the primitive used for solving logistic regression problems in dual form~\cite{fan2008liblinear}. 

For a given problem, the minimum amount of work that needs to be performed by any algorithm is called a {\em lower bound}, and
the algorithms that match the lower bound within a constant factor are called {\em optimal}. 
Parallel algorithms for which the total work performed by all processors is within a constant factor of the state-of-the-art serial algorithm are called {\em work-efficient}.
A parallel algorithm is said to have a {\em data race} whenever multiple threads access the same part of the memory and at least one of those accesses is a write operation.
Whenever there is a data race among threads, the algorithm needs a {\em synchronization} mechanism to avoid inconsistencies.  

In this work, we show that existing shared-memory parallel SpMSpV algorithms are not work-efficient because
they start spending more time accessing the sparse matrix data structure than doing arithmetic as parallelism increases.
We present a new shared-memory parallel SpMSpV algorithm. When the input and output vectors are not sorted, the algorithm is {\em optimal} for matrices with
at least one nonzero per column on average {\em work-efficient}. 
The key insight is to avoid each thread individually scan the list of matrix columns, which is unscalable even if the columns are stored in a sparse format. 
We also implement and evaluate a variation of our algorithm where the input and output vectors are sorted, as it shows better performance in practice due to increased 
cache efficiency. Both variations of the algorithm avoid unnecessary {\em synchronization}.  We experimentally evaluate our algorithm on a Intel Ivy Bridge multicore processor as well
as the new Intel Knight's Landing processor on a variety of real-world matrices with varying nonzeros structures and topologies.

\section{Background}
\subsection{Notation}

Sparse matrix-sparse vector multiplication is the operation $\mathbf{y} \gets \mA \mathbf{x}$ where a sparse matrix $\mA \in \mathbb{R}^{m\times n}$ is multiplied by a sparse
vector $\mathbf{x} \in \mathbb{R}^{n\times 1}$ to produce a sparse vector $\mathbf{y} \in \mathbb{R}^{m\times 1}$. Intuitively, a matrix (vector) is said to be sparse when it is
computationally advantageous to treat it differently from a dense matrix (vector). 
In this paper, we only consider this ``right multiplication'' with the column vector case because the ``left multiplication'' $\mathbf{y'} \gets \mathbf{x'} \mA$ by the row vector 
$\mathbf{x'}$ is symmetric and the algorithms we present can be trivially adopted to the ``left multiplication'' case.

The $\dnnz()$ function computes the number of nonzeros in its input, e.g.,  $\dnnz(\mathbf{x})$ returns the number of nonzeros in  $\mathbf{x}$.
The $\dnzc()$ function, which is only applicable to matrices, computes the number of nonempty columns of its input. When the object is clear from the 
context, we sometimes drop the input and simply write $\dnnz$ and $\dnzc$. We follow
the Matlab colon notation: $\mA(:,i)$ denotes the $i$\dth\ column, $\mA(i,:)$ denotes  the $i$\dth\ row, and $\mA(i,j)$ denotes the 
element at the $(i,j)$\dth\ position of matrix $\mA$.

Our SpMSpV algorithm works for all inputs with different sparsity structures as evidenced by our experimental results, but 
we will analyze its computational complexity on \erdosrenyi random graphs for simplicity. 
In the \erdosrenyi random graph model $G(n,p)$,  each edge is present with probability $p$ independently from each other. For $p=d/m$ where $d \ll m$, in expectation 
$d$ nonzeros are uniformly distributed in each column.  
We use $f$ as shorthand of $\dnnz(\mathbf{x})$ in our analysis. 

\subsection{Classes of SpMSpV algorithms}
SpMSpV algorithms can be broadly classified into two classes: {\em vector-driven} and {\em matrix-driven} algorithms. In vector-driven algorithms, the 
nonzeros in the input vector $\mathbf{x}$ drives the computation and data accesses. By contrast, the nonzeros of the matrix $\mA$ drive the computation in matrix-driven algorithms.
In some sense, vector-driven algorithms can be classified as pull-based since the entries of the matrix are selectively pulled depending on the location of the nonzeros in the input vector.
Following the same logic, matrix-driven algorithms can be classified as push-based.
In the vector-driven formulation, the SpMSpV problem becomes reminiscent of merging multiple lists (i.e., scaled columns of $\mA(:,i)$ for which $\mathbf{x}(i) \neq 0$).

\subsection{Sparse matrix and vector data structures}
There is no shortage of sparse matrix formats, most of which were exclusively invented for the sparse matrix-dense vector multiplication (SpMV) operation.
A recent paper includes an up-to-date list of sparse matrix storage formats~\cite{langr2016evaluation}. The SpMV operation can be implemented
by sequentially iterating over the nonzeros of the sparse matrix, hence does not require fast random access to the columns of a matrix. In SpMSpV, however,
only those columns $\mA(:,i)$ for which $\mathbf{x}(i) \neq 0$ needs to be accessed. Consequently, we only consider the storage formats that allow fast random access
to columns. 

The Compressed Sparse Columns (CSC) format is perhaps the most widely used sparse matrix storage format, together with its row analog; the Compressed Sparse Rows. 
CSC has three arrays: \id{colptrs} is an integer array of length $n+1$ that effectively stores pointers to the start and end positions of the nonzeros for each column,
\id{rowids} is an integer array of length $\dnnz$ that stores the row ids for nonzeros, and \id{values} is an array of length $\dnnz$ that stores the numerical values for nonzeros.  
CSC supports random access to the start of a column in constant time. Some implementations of CSC keep the row ids of nonzeros within each column sorted, e.g.
the range $\id{rowids}(\id{colptrs}(i) \ldots \id{colptrs}(i+1))$ is sorted for all $i$, but this is not a universal requirement.

The Double-Compressed Sparse Column (DCSC) format~\cite{ipdps08} further compresses CSC by removing repetitions in the \id{colptrs} array, which arise from empty columns.
In DCSC, only columns that have at least one nonzero are represented, together with their column indices. DCSC requires $O(\dnzc+\dnnz)$ space compared to CSC's $O(n+\dnnz)$.
DCSC can be augmented to support fast column indexing by building an auxiliary index array that enables random access to the start of a column in expected constant time.
This additional array does not increase the asymptotic storage.

There are two commonly utilized methods to store sparse vectors. The {\em list format} simply stores the vector compactly as a list of (index,value) pairs. The list can be sorted
or unsorted. In contrast to its name, 
the actual data structure is often an array of pairs for maximizing cache performance. This format is space efficient, requiring only $O(\dnnz)$ space. It is often 
the format of choice for vector-driven algorithms but inefficient for matrix-driven algorithms because it does not support constant time random access for a given index. 
The alternative {\em bitvector format}~\cite{sundaram2015graphmat} is composed of a $O(n)$-length bitmap that signals whether or not a particular index is nonzero, and 
an $O(\dnnz)$ list of values.

We require our algorithm to produce the output vector $\mathbf{y}$ in the same format that it received the input vector $\mathbf{x}$. For example, if the input is presented in sorted
list format, then the output should also be in sorted list format. This is necessary to ensure the 
usability of our algorithms as part of a larger computational process where the output of one SpMSpV can then be reused as the input of another SpMSpV. 

\subsection{A Lower Bound for SpMSpV}
We present a simple lower bound for multiplying an $m$-by-$n$ matrix that represents 
the \erdosrenyi graph $G(n, d/m)$ by a sparse $n$-by-$1$ vector with $f$ nonzeros.
Since $f = \dnnz(\mathbf{x})$, any SpMSpV algorithm has to access the nonzero entries in $f$ columns of $\mA$.
Since each column of $\mA$ has $d$ nonzeros in expectation, the asymptotic lower bound of SpMSpV is $\Omega(d f)$.

This lower bound assumes no restrictions for storing the matrix and the vectors.
The algorithm we present in this paper attains this lower bound using unsorted vectors. 
No known algorithm attains this lower bound if we require the vectors to be sorted.

\begin{table*}[htbp]
   \centering
   \caption{Classification of parallel SpMSpV algorithms. $t$ denotes the number of threads. SpMSpV-bucket is presented in this paper. \label{tab:existing}}
      \begin{tabular}{@{} l l l l l l  l r @{}} 
      \toprule
      Class & Algorithms &\multicolumn{2}{c}{Data structures} & Merging & Sequential  & Parallelization & Parallel\\
       & &matrix & vector & strategy & complexity  & strategy & complexity\\
      \midrule
            matrix-driven & GraphMat~\cite{sundaram2015graphmat} & DCSC & bitvector & SPA & $O(\dnzc+df)$ &  row-split matrix  and private SPA &  $O(\dnzc+df/t)$ \\
      vector-driven & CombBLAS-SPA~\cite{bfs:11} & DCSC & list & SPA & $O(df)$ &   row-split matrix and private SPA &  $O(f+df/t)$ \\
      vector-driven & CombBLAS-heap~\cite{bfs:11} & DCSC & list & heap & $O(df\lg{f})$ &   row-split matrix and private heap &  $O(df/t\lg{f})$ \\
      vector-driven & SpMSpV-sort~\cite{yang2015fast} & CSC & list & sorting & $O(df\lg{df})$ &   concatenate, sort and prune &  $--$ \\
       vector-driven & SpMSpV-bucket & CSC & list & buckets & $O(df)$ &   2-step merging and private SPA &  $O(df/t)$ \\
      \bottomrule
   \end{tabular}
   \label{tab:booktabs}
\end{table*}

\subsection{Prior work on parallel SpMSpV algorithms}
\label{sec:prior-work}

A summary of existing algorithms are shown in Table~\ref{tab:existing}, with references to their first appearances in the literature. 
Some libraries later extended their algorithms to support each others' ideas as well, but 
Table~\ref{tab:existing} only refers to the first implementation reported. 

Combinatorial BLAS (CombBLAS)~\cite{bulucc2011combinatorial} includes implementations of a variety of vector-driven algorithms. 
The algorithms that use the DCSC format 
has been first used in the context of parallel breadth-first search (BFS)~\cite{bfs:11}. For shared-memory parallelization, the BFS work advocated splitting the matrix
row-wise to $t$ (number of threads) pieces. Each thread local $m/t$-by-$n$ submatrix was then stored in DCSC format. The authors experimented with multiple data structures 
for merging scaled columns of $\mA$: a sparse accumulator (SPA) and a priority queue (heap). The SPA~\cite{gilbert1992sparse} is an abstract data type that (at minimum) consists 
of a dense vector of numerical values and a list of indices that refer to nonzero entries in the dense vector. CombBLAS later extended its support to CSC.

GraphMat~\cite{sundaram2015graphmat} supports a matrix-driven SpMSpV algorithm. In GraphMat, the matrix is represented in the DCSC format and the vector
is stored using the bitvector format. GraphMat also splits matrix row-wise. 
Nurvitadhi et al.~\cite{nurvitadhi2016hardware} present a hardware accelarator for a vector-driven SpMSpV algorithm. 
Their algorithm description makes random accesses to vector $\mathbf{y}$, without any reference to its sparsity. Yang et al.~\cite{yang2015fast} present a vector-driven SpMSpV implementation on the GPU, using sorted input/output vectors.

\subsection{Requirements for a parallel work-efficient SpMSpV algorithm}
\begin{itemize}
\item {\bf To attain the lower bound, an SpMSpV algorithm must be vector driven.} In contrast to the matrix-driven algorithms that need to iterate over $n$ or $nzc$ columns for DCSC and CSC formats, respectively, the vector-driven algorithms can efficiently access $\id{df}$ entries of the matrix. 

\item {\bf To attain the lower bound, a SPA-based SpMSpV algorithm should not initialize the entire SPA.} Since SPA is a dense vector of size $m$, initializing the entire SPA requires $O(m)$ time.  By contrast, an algorithm that only initializes entries of SPA to be accessed in the multiplication requires $O(\dnnz(\mathbf{y}))$ initialization time; hence can be work efficient. 

\item {\bf A parallel SpMSpV algorithm that splits the matrix row-wise  is not work efficient.} Consider an algorithm that splits $\mA$ row-wise to $t$ pieces and multiplies each of the  $m/t$-by-$n$ submatrices independently with $\mathbf{x}$ in parallel by $t$ threads to produce $1/t$ piece of $\mathbf{y}$. Here, each thread needs to access the entire input vector $\mathbf{x}$ requiring $O(f)$ time per thread to access $\mathbf{x}$. The total time to access $\mathbf{x}$ over all  threads is $O(tf)$, making it work inefficient.
However, in the row split case, each thread writes to a separate part of the output vector $\mathbf{y}$, so no synchronization is needed. 

\item {\bf A parallel SpMSpV algorithm that splits the matrix column-wise needs synchronization among threads.} Consider an algorithm that splits $\mA$ column-wise to $t$ pieces and multiplies each of the  $m$-by-$n/t$ submatrices  with $1/t$ piece of $\mathbf{x}$ in parallel by $t$ threads to produce  $\mathbf{y}$. 
This algorithm is work efficient because the nonzero entries of $\mathbf{x}$ and $\mA$ are accessed only once. 
However, synchronization is required among threads in the column split case because each thread writes to the same output vector $\mathbf{y}$ via a shared SPA. 

\item {\bf A parallel SpMSpV algorithm that employs 2-D partitioning of the matrix is not work efficient.} Consider an algorithm that partitions $\mA$ into $\sqrt{t}\times \sqrt{t}$ grids and multiplies each of the  $m/\sqrt{t}$-by-$n/\sqrt{t}$ submatrices with $1/\sqrt{t}$ piece of $\mathbf{x}$ to generate partial $1/\sqrt{t}$ piece of $\mathbf{y}$.
Since each submatrix in a column of the grid needs to access the same $1/\sqrt{t}$ piece of $\mathbf{x}$, the input vector is accessed $\sqrt{t}$ times across all threads, making the algorithm work inefficient.
Futhermore, threads processing submatrices in a row of the grid need to update the same part of the output vector $\mathbf{y}$, requiring synchronization among threads. 
This algorithm mimics the concepts of distributed-memory SpMSpV algorithms in CombBLAS and GraphMat.

\end{itemize}

\begin{table}[htbp]
   \centering
     \caption{Characteristics of SPA-based sequential and parallel SpMSpV algorithms. $^1$ In column-split and 2-D partitioning based algorithms, private SPA is not considered because it requires $O(tm)$ memory for $t$ threads.}
   \begin{tabular}{@{} ll c c c @{}} 
         \toprule
         & Algorithm & Attain  &  Work   &  Synch. \\
         & aspects & lower bound? &  efficient?  &  needed?\\
          \toprule
      \parbox[t]{2mm}{\multirow{4}{*}{\rotatebox[origin=c]{90}{Sequential}}} & matrix driven    & $\times$  \\
      & vector driven & \checkmark\\
      & SPA full init & $\times$  \\
      & SPA partial init & \checkmark \\
       \midrule
     \parbox[t]{2mm}{\multirow{3}{*}{\rotatebox[origin=c]{90}{Parallel}}}  & row-split (private SPA) & & $\times$ & $\times$ \\
       & column-split (shared SPA$^1$) &  & \checkmark & \checkmark \\
      & 2-D (shared SPA$^1$) &  & $\times$ & \checkmark\\

      \bottomrule
   \end{tabular}

   \label{tab:summary-SPA}
\end{table}

We summarize the properties SPA-based sequential and parallel SpMSpV algorithms in Table~\ref{tab:summary-SPA}.
Based on this table, an asymptotically optimal SpMSpV algorithm that attains the lower bound should be vector-driven and initializes only necessary entries of SPA.
A desirable parallel algorithm should be work-efficient and should perform as little synchronization as possible (synchronization-avoiding).
However, none of the parallelization scheme described in Table~\ref{tab:summary-SPA} is both work-efficient and synchronization-free at the same time. 
This observation motivates us to develop a new parallel algorithm incorporating the advantages of both row- and column-split schemes to make the newly-developed 
algorithm both work efficient and synchronization-free.
In contrast to CombBLAS and GraphMat that split the matrix row-wise beforehand, our algorithm, called SpMSpV-bucket, splits the necessary columns of the matrix on the fly 
using a list of buckets.
This approach can address the need of each multiplication independently and has been shown to be very effective in sparse matrix-dense vector multiplication~\cite{buono2016optimizing}.
We describe the SpMSpV-bucket algorithm in the next section.

\section{The SpMSpV-bucket algorithm}
\label{sec:SpMSpV-bucket}

\begin{algorithm}[htbp]
\begin{algorithmic}[1]
\begin{small}
\Procedure{SpMSpV}{$\mA$, $\mathbf{x}$, \id{SPA}, \id{Buckets}}
\LineComment{\textcolor{blue}{Step1: Gather necessary columns of $\mA$ in $t$ buckets (each bucket corresponds to a subset of consecutive rows of the matrix)}}
\For {every nonzero entry $(j,x(j))$ in $\mathbf{x}$} {\bf in parallel }
	\For {every nonzero $\mA(i,j)$ in $\mA(:,j)$ }  
		\State $\id{k} \gets \floor{(i \times nb) / m}$ \Comment{the destination bucket}
		\begin{tcolorbox}[width=\linewidth, boxsep=0pt, left=-.5pt, right=0pt, top=1pt, bottom=1pt, boxrule=0.5pt, colback=red!5!white,colframe=red!75!black]
		 \LineComment{Lock-free insertion, see text for details}
		\State $B_k \gets B_k \cup (i, \Call{Mult}{\mathbf{x}(j),\mA(i,j)})$
		\end{tcolorbox} 
	\EndFor 
\EndFor 	
\For{each bucket $B_k$ in \id{Buckets}} {\bf in parallel }

	\State $uind_k \gets \phi$  \Comment{unique indices found in this bucket}
	\LineComment{\textcolor{blue}{Step2: Merge entries in each bucket}}
	\For{every ($ind, val$) pair in \id{B_k}}  
		\State $\id{SPA}[ind] \gets \infty$
	\EndFor

	\For{every ($ind, val$) pair in \id{B_k}}  
		\If {$\id{SPA}[ind] = \infty$}
                    	\State $\id{uind_k} \gets uind_k  \cup ind$ \Comment{save unique indices}
			\State $\id{SPA}[ind] \gets val$
		\Else
			\State $\id{SPA}[ind] \gets \Call{Add}{\id{SPA}[ind] , val}$
		\EndIf
	\EndFor
	
	\LineComment{\textcolor{blue}{Step3: Construct $\mathbf{y}$ by concatenating buckets using SPA}}
	\State $\id{offset_k} \gets \sum_{l=0}^{k-1} |\id{uind_l}| $ \Comment{using prefix sum on the master thread} 
	\State $i \gets 0$
		\For{each $ind$ in \id{uind_k}} 
			\State $y[\id{offset_k}+i] \gets (ind, \id{SPA}[ind])$ 
			\State $i \gets i + 1$
		\EndFor
\EndFor	

 \EndProcedure
\end{small}
\end{algorithmic}
\caption{Parallel SpMSpV algorithm. {\bf Input:}  An ${m\times n}$ sparse matrix $\mA$ stored in CSC format, the input sparse vector $\mathbf{x}$, a dense vector \id{SPA} of size $m$, and a list of $nb$ buckets \id{Buckets}.  {\bf Output:} the output sparse vector $\mathbf{y}$.}
\label{alg:SpMSpV}
\end{algorithm}

\begin{algorithm}[!th]
\begin{algorithmic}[1]
\begin{small}
\Procedure{Estimate-Buckets}{$\mA$, $\mathbf{x}$, \id{Buckets}}
\For {$t$ in $1..nt$} {\bf in parallel } \Comment{$nt$ is the number of threads}
	\State $\id{Boffset}[t]\gets 0$  \Comment{initialize to zero}
	\State $\mathbf{x}_t \gets 1/t$ piece of $\mathbf{x}$ processed by the $t$-th thread
	\For {every nonzero entry $(j,\mathbf{x}_t(j))$ in $\mathbf{x}_t$} 
		\For {every nonzero $\mA(i,j)$ in $\mA(:,j)$ }  
			\State $\id{b} \gets \floor{(i \times nb) / m}$ \Comment{destination bucket}
			\State $\id{Boffset}[t][b]\gets \id{Boffset}[t][b]+1$
		\EndFor 
	\EndFor 	
\EndFor 	
 \EndProcedure
\end{small}
\end{algorithmic}
\caption{Preprocessing step of parallel SpMSpV algorithm needed to avoid synchronization among threads when inserting entries to buckets. {\bf Input:}  see Algorithm~\ref{alg:SpMSpV}.  {\bf Output:} An $nt$-by-$nb$ array $\id{Boffset}$ where  \id{Boffset[i][j]} stores the number of entries that the $i$th thread will insert to $j$th bucket in Step 1 of Algorithm~\ref{alg:SpMSpV}.}
\label{alg:Estimate-Buckets}
\end{algorithm}

\begin{figure*}[htbp]
   \centering
   \includegraphics[scale=.55]{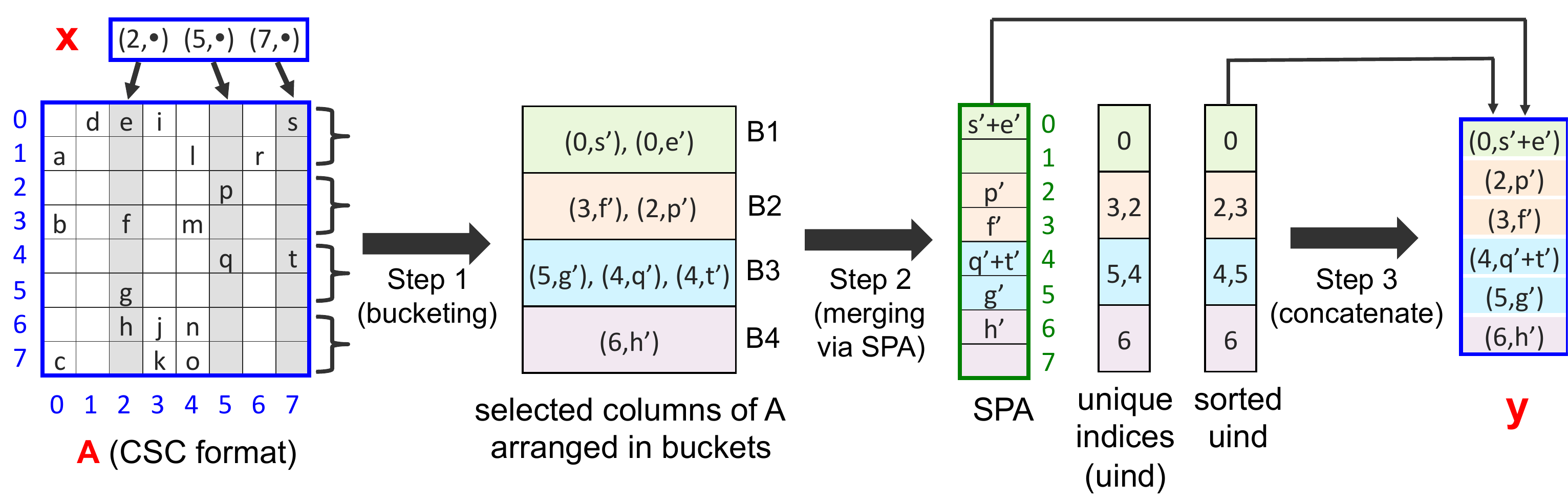} 
   \caption{Three steps of the SpMSpV algorithm. In the first step, nonzero entries of the selected columns of $\mA$ are multiplied by the corresponding elements of $\mathbf{x}$. The multiplied values (denoted with prime symbols) coupled with their row indices are stored in four buckets. The bucket where an entry is stored is determined by its row index. Data structures possessed or touched by four buckets are shown in four different colors. In the second step, entries in each bucket are merged independently by using a sparse accumulator. In each bucket, unique indices (\id{uind}) are identified and sorted (sorting is an optional step and is only performed when sorted output is required or to improve cache locality). In the third step, the output vector $\mathbf{y}$ is created by concatenating \id{uind} from all buckets and fetching the corresponding values from the SPA.}
   \label{fig:example}
\end{figure*}

Algorithm~\ref{alg:SpMSpV} describes the steps of the SpMSpV-bucket algorithm that takes a dense vector \id{SPA} of size $m$ and a list of \id{nb} buckets along with $\mA$ and  $\mathbf{x}$ as inputs.
The matrix is stored in CSC format and the vector is stored in list format.
The buckets are uninitialized space to be used by threads to temporarily store (row index, scaled value) pairs from the  selected columns of the matrix.
Each bucket corresponds to a subset of consecutive rows of the matrix.
The $i$th location of SPA corresponds to the  $i$th row of the matrix and  is accessed by a single thread only.
The SpMSpV-bucket algorithm then performs the following three steps for the multiplication.

{\bf Step 1: Accumulate columns of $\mA$ into buckets (lines 2-7 of Algorithm~\ref{alg:SpMSpV}).} 
In this step, the columns $\mA(:,i)$ for which $\mathbf{x}(i) \neq 0$ are extracted, the values of the extracted columns are multiplied by the nonzero values of $\mathbf{x}$, and  the scaled values paired with their row indices are stored in buckets.   
The bucket in which a scaled matrix entry is placed is determined by its row index. 
More specifically, values in the $i$th row are stored in ($\floor{(i \times nb) / m}$)-th bucket where \id{nb} is the number of buckets.
This step is depicted in Step 1 of Figure~\ref{fig:example} where the second, fifth and seventh columns of $\mA$ corresponding to nonzero indices of $\mathbf{x}$ are extracted and stored in four buckets B1, B2, B3, and B4.
This step is similar to the column-split algorithm that ensures the work-efficiency of our parallel algorithm.

In the parallel algorithm,  each thread processes a subset of nonzero entries of $\mathbf{x}$ and stores the scaled entries of the corresponding columns of $\mA$ in their designated buckets.
Writing to buckets requires synchronization among threads because multiple threads could write simultaneously to the same bucket when they extract entries from the same row $\mA$.
To avoid expensive synchronizations, we pass over the columns of $\mA(:,i)$ for which $\mathbf{x}(i) \neq 0$ in a preprocessing step and count how many scaled entries each thread will write to a bucket in Step 1 of Algorithm~\ref{alg:SpMSpV}.
The preprocessing step is described in Algorithm~\ref{alg:Estimate-Buckets} where \id{Boffset[i][j]} stores the number of entries that the $i$th thread will insert to $j$th bucket in Step 1 of Algorithm~\ref{alg:SpMSpV}.
We use \id{Boffset} to precisely compute where each thread will insert in each bucket.
Using this approach, threads can insert to buckets (line 7 of Algorithm~\ref{alg:SpMSpV}) without any synchronization.

{\bf Step 2: merge entries in each  bucket (lines 10-18 of Algorithm~\ref{alg:SpMSpV}).}
At this point, the algorithm behaves like a row-split algorithm where the buckets store scaled entries split row-wise among the buckets. 
Since there is no data dependency among buckets after they are populated, a bucket can be merged independently by a thread.
At the beginning of this step, each thread initializes only those locations of the SPA to be used in merging entries in the current bucket. 
Next, entries in a bucket are merged using a part of SPA dedicated only for this bucket.
In this process, the algorithm retrieves unique indices from the $k$th bucket and stores them in \id{uind_k}.
This step is depicted in Step 2 of Figure~\ref{fig:example} where each of the four buckets independently  merges its entries by adding values with the same row indices.

{\bf Step 3: Construct $\mathbf{y}$ by concatenating buckets using SPA (lines 19-24 of Algorithm~\ref{alg:SpMSpV}).}
In the final step, unique indices identified in a bucket are coupled with the corresponding values of the SPA and the (index, value)  pairs are inserted to the result vector.
To make this step synchronization free, unique indices in a bucket are mapped to indices of $\mathbf{y}$ using a prefix sum operation described in line 20 of Algorithm~\ref{alg:SpMSpV}.
This step is depicted in Step 3 of Figure~\ref{fig:example} where six unique indices are coupled with the computed values of SPA and saved in the output vector $\mathbf{y}$.
In Figure~\ref{fig:example}, we also showed the situation when indices in $\mathbf{y}$ are required to be sorted.

So far, we have not addressed the sortedness of the vectors. The algorithm works as-is for unsorted vectors. The sortedness of the input $\mathbf{y}$ does not affect the correct 
of the algorithm. However, in order to return a sorted output, the algorithm requires a modification to add a sorting step at the very end.

\subsection{Performance optimizations}
{\bf Load balancing.} In order to balance work among threads, we create more buckets than the available number of threads. 
In our experiments, we use $4t$ buckets when using $t$ threads and employ dynamic scheduling of threads over the buckets.
Using more buckets tends to improve the scalability of the SpMSpV-bucket algorithm except when the input vector is extremely sparse (see the discussion in Section~\ref{sec:perf-breakdown}).

{\bf Cache efficiency.}
To improve the cache locality of Step 1 in Algorithm~\ref{alg:SpMSpV}, we allocate a small private buffer for each thread.
A thread first fills its private buffer as it accesses the columns of $\mA$ and copies data from the private buffer to buckets when the local buffer is full.
The thread-private buffer is small enough to fit in L1 or L2 cache.
Sorting the input vector  $\mathbf{x}$ beforehand (if it is not sorted) improves the cache locality of the bucketing step when $\mathbf{x}$ is denser.
This is due to the fact that when $\mathbf{x}$ is denser, the probability of accessing consecutive columns of $\mA$ increases significantly.

{\bf Memory allocation.}
The memory allocation time for buckets and SPA can be expensive, especially when we run SpMSpV many times in an iterative algorithm such as the BFS.
Hence, we allocate enough memory for all buckets and SPA only once and pass them to the SpMSpV-bucket algorithm.
The number of entries inserted in all buckets is at most $O(\dnnz(\mA))$.
Hence, preallocating the buckets does not increase the total memory requirement of our algorithm.

\subsection{Time and space complexity}
{\bf Serial complexity.} 
The preprocessing step described in Algorithm~\ref{alg:Estimate-Buckets} and  Step 1 in Algorithm~\ref{alg:SpMSpV} both access $df$ nonzero entries from $f$ columns of $\mA$. 
Hence these steps require $O(df)$ time.
The initialization of SPA and merging entries in all buckets require another $O(df)$ time in the second step.
The total number of entries in $\id{uind_k}$ across all buckets is $\dnnz(\mathbf{y})$.
Since $\dnnz(\mathbf{y}) \leq df$, the overall complexity of the algorithm is $O(d f)$.
If $\mathbf{y}$ is needed to be sorted by nonzero indices, another $O(\dnnz(\mathbf{y}) \log{\dnnz(\mathbf{y}))}$ time is required for sorting.
However, sorting is very efficient in SpMSpV-Bucket algorithm because only unique indices in each buckets are needed to be sorted.
Hence each thread can run a sequential integer sorting function on its local indices using efficient sorting algorithms such as the radix sort.  

{\bf Parallel complexity.} In the first step, $f$ nonzero entries of the input vector are evenly distributed among $t$ threads.
Hence, each thread accesses $fd/t$ nonzero entries of the matrix.
Since the nonzero entries of the matrix are evenly distributed among rows in the \erdosrenyi model, each bucket will have $fd/t$ entries in
expectation when $t$ buckets are used.
Hence the parallel complexity of the SpMSpV-Bucket algorithm is $O(fd/t)$.

{\bf Space complexity.} The total space required for all buckets is no more than $O(\dnnz(\mA))$.
Hence total space requirement of our algorithm is  $O(m + \dnnz(\mA))$.

{
\setlength{\tabcolsep}{5pt}
\begin{table}[!t]{
\centering
\begin{tabular}{rcc}
					&  {\bf Cori} & {\bf Edison }    \\
					& ({\bf Intel KNL})  & ({\bf Intel Ivy Bridge})		\\
\hline
{\bf Core }	 	 & 			& \\

\hline
Clock (GHz)			& 1.4			& 2.4					\\
L1 Cache (KB)		& 32		& 32				\\
L2 Cache (KB)		& 1024$^1$		& 256				\\
DP GFlop/s/core		& 44			&19.2		\\
\hline
{\bf Node Arch.}	 	 & 			& \\
\hline
Sockets/node			&  1		&	2					\\
Cores per socket			& 64				& 12					\\
STREAM BW$^2$		&  102~GB/s 	&	104~GB/s		\\
Memory per node		&  96~GB	&	64~GB			\\
\hline
{\bf Prog. Environment}	 	 & 			& \\
\hline
Compiler & gcc 5.3.0 & gcc 5.3.0\\
Optimization & -O3 &  -O3 \\
\hline
\end{tabular}

\caption{Overview of Evaluated Platforms.  $^1$Shared between 2 cores in a tile. $^2$Memory bandwidth is measured using the STREAM copy benchmark per node.}
\label{tab:machines}
}
\end{table}
}

\begin{table*}[!ht]
 \centering
 \caption{Test problems from the University of Florida sparse matrix collection~\cite{davis2011university}. 
}

 \begin{tabular}{@{} l l c  c   c   l @{}}
    \toprule
    
 Class & Graph	&	\#vertices	&	\#edges &	 pseudo & Description\\
 & &	($\times 10^6$)	&	($\times 10^6$) 	&	diameter  &	\\

  \toprule
& amazon0312	&	0.40	&	3.20			& 21	&	Amazon product co-purchasing network	\\
& web-Google	&	0.92	&	5.11			&	16	&	Webgraph from the Google prog. contest, 2002	\\
low-diameter graphs  &  	wikipedia-20070206	&	3.56	&		45.03	&	14	&	Wikipedia page links	\\
 & ljournal-2008	&	5.36	&		79.02 & 34	&	LiveJournal social network\\
  &  wb-edu	&	9.85	&	57.16	&	38	&	Web crawl on .edu domain	\\
    &  dielFilterV3real	&	1.10	&		89.31	&	84 &		High-order vector finite element method in EM\\

  \midrule 
  &  G3\_circuit	&	1.56	&	7.66	&	514 &		circuit simulation problem	\\ 
   &   hugetric-00020	&	7.12		&	21.36		&	3,662 &		undirected graph\\	
  high-diameter graphs&   hugetrace-00020	&	16.00		&	48.00		&	5,633 &		Frames from 2D Dynamic Simulations\\
 &  delaunay\_n24	&	16.77		&	100.66	& 1,718	&		Delaunay triangulations of random points\\ 

 & rgg\_n24\_s0	&	16.77	&	165.1	&	3,069  &	Random geometric graph\\ 

  	       \toprule
  \end{tabular}
\label{table:problem-statistics}
 \end{table*}

\section{Results}
\label{sec:results}

\subsection{Experimental Setup} 
We evaluate the performance of SpMSpV algorithms on Edison, a Cray XC30 supercomputer at NERSC and on a KNL manycore porcessor that will be integrated with NERSC/Cori.
These two systems are described in Table~\ref{tab:machines}.
We used OpenMP for multithreaded execution in our code. 

 Table~\ref{table:problem-statistics} describes a set of real matrices from the University of Florida sparse matrix collection~\cite{davis2011university} used in our experiments.
 We selected the low-diameter scale-free graphs and high-diameter graphs arising in various scientific domains.
%

\subsection{Impact of sorted input and output vectors on the performance of the SpMSpV-bucket algorithm}
\begin{figure*}[!t]
   \centering
   \includegraphics[scale=.8]{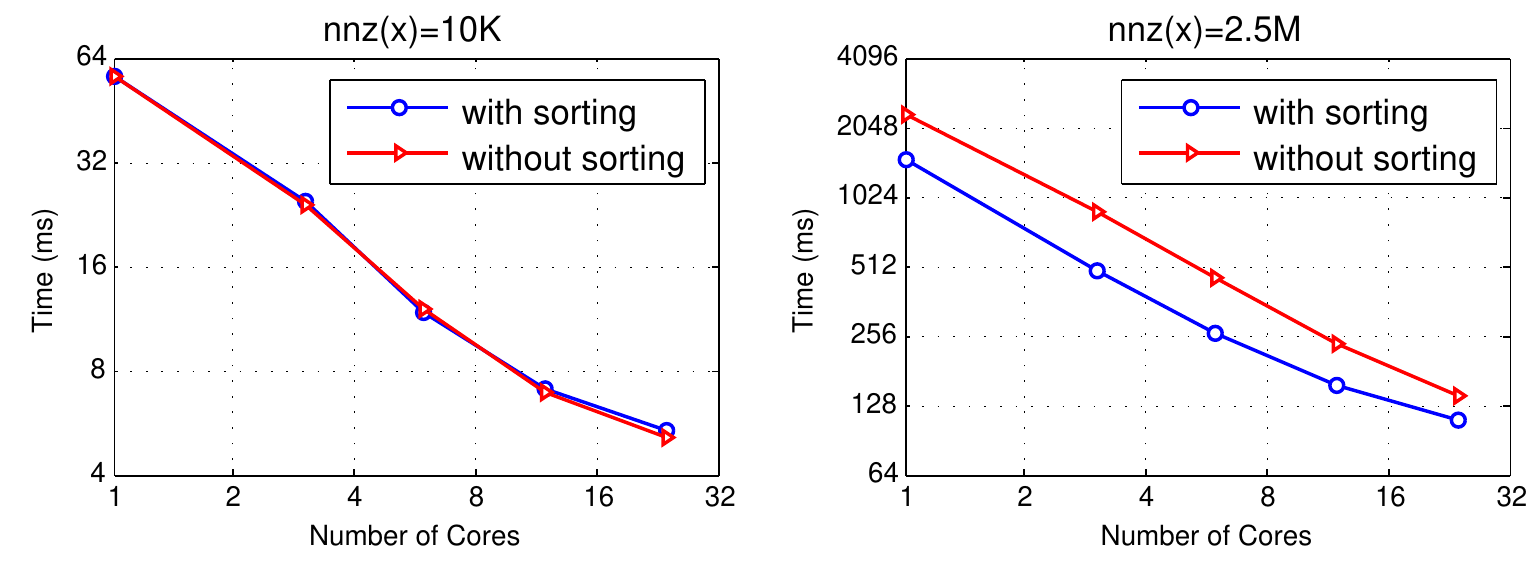} 
   \caption{Runtime of the SpMSpV-bucket algorithm with or without sorted input and output vectors. Here, the adjacency matrix of \texttt{ljournal-2008} is multiplied by sparse vectors with (a) 10K and (b) 2.5M nonzeros. The experiment was run on Edison. }
   \label{fig:sorting}
\end{figure*}

We implemented two variants of the SpMSpV-bucket algorithm based on the sortedness of the input and output vectors: in one variant both $\mathbf{x}$ and $\mathbf{y}$ are kept sorted by their indices, while the second variant works on unsorted vectors.
Figure~\ref{fig:sorting} shows the impact of sorted vectors on the performance of the SpMSpV-bucket algorithm for $\mathbf{x}$ with 10K and 2.5M nonzeros.
When the vector is relatively dense, keeping the vectors sorted improves the performance of our algorithm as can be seen in the right subfigure in  Figure~\ref{fig:sorting}.
This is due to the fact that when $\mathbf{x}$ is denser, the probability of accessing consecutive columns of $\mA$ increases, making the bucketing step (Step 1 in Algorithm 1) more cache efficient.
However, for relatively sparse vectors (e.g., when $\dnnz(\mathbf{x})$ is less than $1\%$ of $n$), sorted vectors do not significantly impact the performance of the SpMSpV-bucket algorithm because the access pattern to columns of $\mA$ more or less remains random.
Since the unsorted version never seems to outperform the sorted version in practice, we only present results with sorted vectors in the remainder of the results section.

\begin{figure*}[!t]
   \centering
   \includegraphics[scale=1.1]{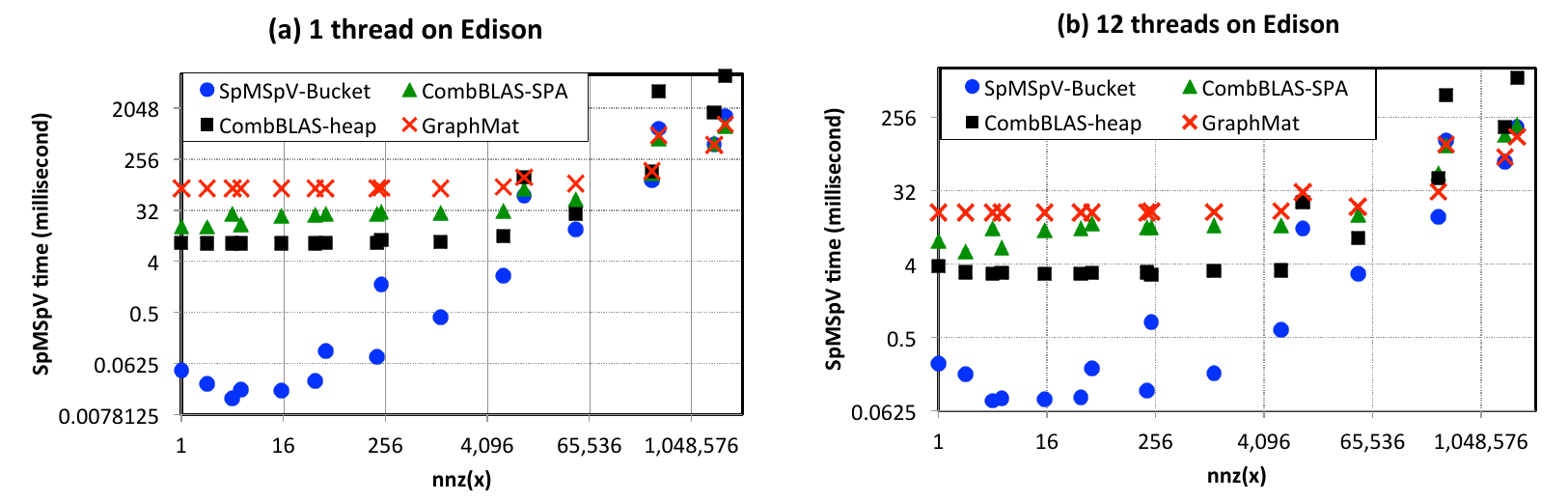} 
   \caption{Runtime of four SpMSpV algorithms when the adjacency matrix of \texttt{ljournal-2008} is multiplied by sparse vectors with different number of nonzero entries using (a) 1 thread and (b) 12 threads on Edison. The sparse vectors represent frontiers in a BFS starting from the first vertex of \texttt{ljournal-2008}. }
   \label{fig:multiply_nnzx}
\end{figure*}

\subsection{Relative performance of SpMSpV algorithms}
\label{sec:rel-perf}
We compare the performance SpMSpV-bucket with three other SpMSpV algorithms: CombBLAS-SPA, CombBLAS-heap, and GraphMat.
These algorithms are already discussed in Section~\ref{sec:prior-work}.
At first, we study the impact of $\dnnz$ of the input vector $\mathbf{x}$ on the performance of  SpMSpV algorithms.
Figure~\ref{fig:multiply_nnzx} shows the runtime of four algorithms when the adjacency matrix of \texttt{ljournal-2008} is multiplied by sparse vectors with values of $\dnnz(\mathbf{x})$  using (a) 1 thread and (b) 12 threads on Edison.
When $\mathbf{x}$ is very sparse (i.e., $\dnnz(\mathbf{x})$ less than 50K), the runtime of GraphMat remains constant for a fixed thread count.
This is a property of any matrix-driven algorithm whose runtime is dominated by the $O(\dnzc)$ term needed to pass through all nonzero columns of the matrix,  especially when the vector is very sparse.
CombBLAS-SPA also shows similar behavior for very sparse vectors because it fully initializes the SPA requiring $O(m)$ time.
By contrast, SpMSpV-bucket does not have any extra overhead when the vector is very sparse; hence it outperforms its competitors by several orders of magnitude. 
For example, when $\dnnz(\mathbf{x})=50$, SpMSpV-bucket  is $200\times$, $81\times$, and $744\times$ faster than CombBLAS-SPA, CombBLAS-heap, and GraphMat, respectively on a single thread.
When $\dnnz(\mathbf{x})=1100$, SpMSpV-bucket  is $68\times$, $21\times$, and $191\times$ faster than CombBLAS-SPA, CombBLAS-heap, and GraphMat, respectively on a single thread.
This huge gap in performance shrinks as the input vector becomes denser when $O(\dnzc)$ terms is less significant.
For example, when $\dnnz(\mathbf{x})=1.9M$, SpMSpV-bucket, CombBLAS-SPA, and GraphMat all performs similarly and run $3.5\times$ faster than CombBLAS-heap because of the logarithm term in the latter algorithm.
These story remains more or less similar on higher concurrency as can be seen in Figure~\ref{fig:multiply_nnzx}(b).


\begin{figure*}[!t]
   \centering
   \includegraphics[scale=.95]{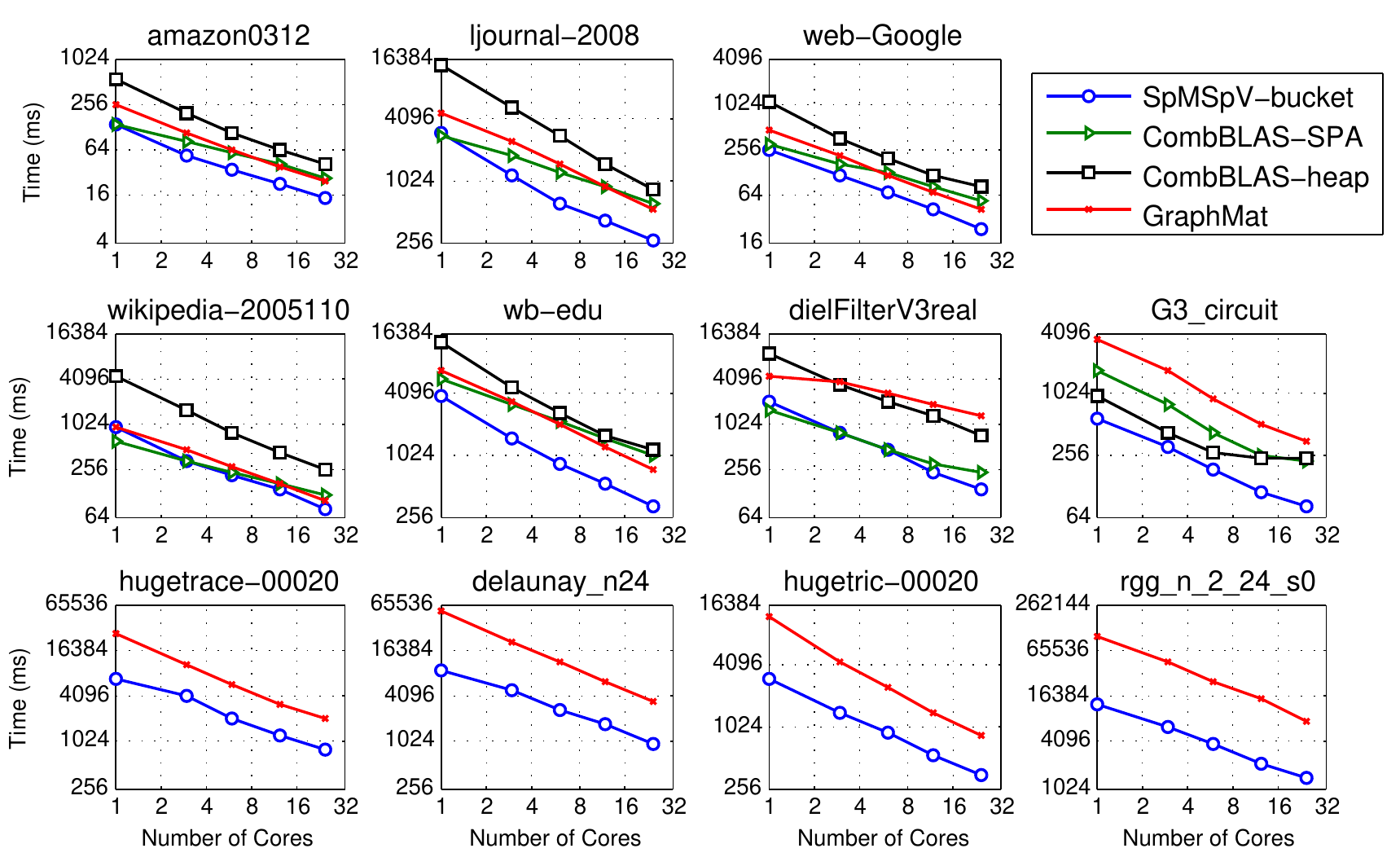} 
   \caption{Strong scaling of four shared-memory SpMSpV algorithms when they are used in BFS. The experiments were run on a single node of Edison. For each graph, the same source vertex is used to start the BFS by all four algorithms. We only report the runtime of SpMSpVs in all iterations omitting other costs of the BFS. For the high-diameter graphs in the bottom row, CombBLAS-DCSC and heap-merge algorithms were not competitive, hence we omit them for these graphs.}
   \label{fig:bfs-SpMSpV-edison}
\end{figure*}

\begin{figure*}[!t]
   \centering
   \includegraphics[scale=.85]{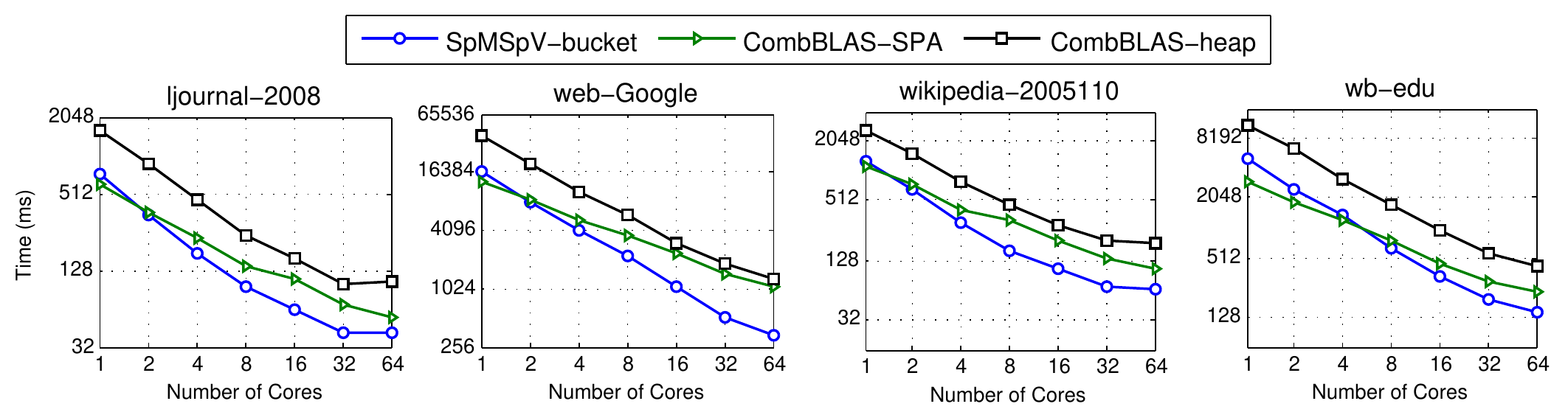} 
   \caption{Strong scaling of three shared-memory SpMSpV algorithms when they are used in BFS on KNL. For each graph, the same source vertex is used to start the BFS by all four algorithms. We only report the runtime of SpMSpVs in all iterations omitting other costs of the BFS. We were unable to run GraphMat on KNL.}
   \label{fig:bfs-SpMSpV-KNL}
\end{figure*}

\begin{figure*}[!t]
   \centering
   \includegraphics[scale=.98]{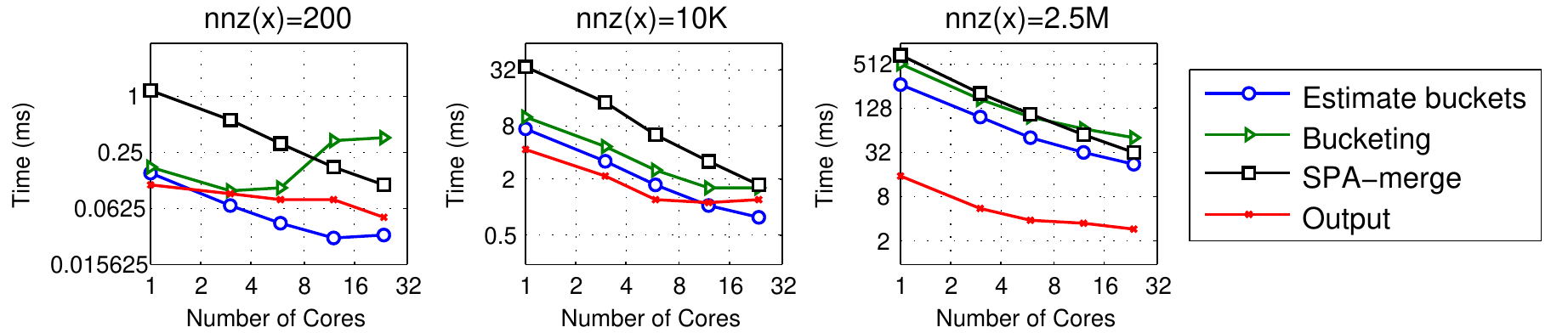} 
   \caption{Strong scaling of four components of SpMSpV-bucket algorithm when the adjacency matrix of \texttt{ljournal-2008} is multiplied by sparse vectors with different number of nonzeros on Edison.}
   \label{fig:time-breakdown-scaling}
\end{figure*}

\subsection{Performance of SpMSpV algorithms when used in BFS}
BFS is arguably the most common customer of SpMSpV where the product of  the adjacency matrix of the graph and the sparse vector representation of the current frontier provides the next frontier of the BFS.
This approach has been successfully used in parallel BFS targeting GPU and the shared- and distributed-memory platforms~\cite{bfs:11, sundaram2015graphmat, yang2015fast}. 
Here we compare the performance for four SpMSpV algorithms when they are used in BFS.

Figure~\ref{fig:bfs-SpMSpV-edison} shows the performance of four shared-memory SpMSpV algorithms on eleven real world matrices from Table~\ref{table:problem-statistics} on a single node of Edison.
To ensure the fairness in comparing algorithms, the same source vertex is used to start the BFS by all four algorithms and only the runtime of SpMSpVs in all iterations are considered.
For all problems in Figure~\ref{fig:bfs-SpMSpV-edison}, SpMSpV-bucket runs the fastest for all concurrencies. 
The performance improvement is more dramatic on high-diameter graphs where SpMSpV-bucket runs ${3\times}$ to  $10\times$ faster than GraphMat as can be seen in the bottom row of Figure~\ref{fig:bfs-SpMSpV-edison}.
According to the discussion in Section~\ref{sec:rel-perf}, this performance gap is expected for high-diameter graphs where BFS executes many SpMSpVs with very sparse vectors -- a territory where matrix-driven algorithms are inefficient.
On scale-free graphs, SpMSpV-bucket  still performs the best, but the gaps among the algorithms are narrower.
This is due to the fact that BFS on a scale-free graph is usually dominated by few iterations with dense frontiers where matrix-driven algorithms usually perform their best. 

On average, SpMSpV-bucket achieves $11\times$ (max: $14\times$, min: $9\times$), CombBLAS-SPA achieves $6\times$ (max: $7\times$, min: $5\times$), CombBLAS-heap achieves 
$12\times$ (max: $17\times$, min: $4\times$), and GraphMat achieves, $11\times$ (max: $15\times$, min: $9\times$) speedups, when going from 1 thread to 24 threads on Edison.
GraphMat attains better scalability, even when $\mathbf{x}$ is very sparse, because  it always has $O(\dnzc)$ work needed to access nonzero columns of the matrix.
By contrast, our work-efficient algorithm might not scale well when the vector is very sparse (e.g., $\dnnz(\mathbf{x})$ less than the number of threads) due to the lack of enough work for all threads.
The parallel efficiency of CombBLAS-SPA decreases with increasing concurrency because the total amount of work performed by all threads increases as each 
thread scans the entire input vector.
Poor serial runtime often contributes to the high speedups of the CombBLAS-heap algorithm.

\subsection{Performance of SpMSpV algorithms on the Intel KNL processor}
Figure~\ref{fig:bfs-SpMSpV-KNL} shows the performance of three SpMSpV algorithms on the Intel KNL processor equipped with 64 cores.
We were unable to run GraphMat on KNL.
On average, SpMSpV-bucket achieves $32\times$ (max: $49\times$, min: $20\times$), CombBLAS-SPA achieves $12\times$ (max: $14\times$, min: $10\times$), and CombBLAS-heap 
achieves $20\times$ (max: $30\times$, min: $12\times$) speed-up when going from 1 thread to 64 threads on KNL.
As before, the serial performance of CombBLAS-SPA is similar to or slightly better than SpMSpV-bucket on scale-free graphs. 
However, scalability of CombBLAS-SPA suffers with increasing number of threads because of its work inefficiency.
By contrast, SpMSpV-bucket scales well up to 64 cores of KNL for diverse classes of matrices. 
We did not observe any benefit of using multiple threads per core on KNL.

\subsection{Performance breakdown of the SpMSpV-bucket algorithm}
\label{sec:perf-breakdown}
The SpMSpV-bucket algorithm has four distinct steps (including the preprocessing step) that are described in Section~\ref{sec:SpMSpV-bucket}.
Here we show how these steps contribute to the total runtime of SpMSpV and how they scale as we increase thread count.
Figure~\ref{fig:time-breakdown-scaling} shows the strong scaling of the components of the SpMSpV-bucket algorithm when the adjacency matrix of \texttt{ljournal-2008} is multiplied by $\mathbf{x}$ with different sparsity patterns.
SPA-based merging is the most expensive step of the sequential SpMSpV-bucket algorithm for all sparsity patterns of the input vector. As  $\mathbf{x}$ becomes denser, bucketing becomes as expensive as merging on a single thread.
For example, in Figure~\ref{fig:time-breakdown-scaling}, SPA-based merging takes 73\%, 62\%, and 46\% of the total sequential runtime when $\dnnz(\mathbf{x})$ is 200, 10K and 2.5M, respectively. By contrast, the bucketing steps takes 10\%, 17\%, and 35\% of the serial runtime  when $\dnnz(\mathbf{x})$ is 200, 10K and 2.5M, respectively.

SPA-based merging has the best scalability than other steps of the SpMSpV-bucket algorithm for all sparsity levels of $\mathbf{x}$ because each thread independently performs 
the merging on its private bucket.
For example, when we go from 1 core to 24 cores in Figure~\ref{fig:time-breakdown-scaling}, SPA-based merging achieves $11\times$, $19\times$, and $22\times$ speedups when $\dnnz(\mathbf{x})$ is 200, 10K and 2.5M, respectively.
By contrast, the bucketing step achieves $6\times$, and $10\times$ speedups  when $\dnnz(\mathbf{x})$ is 10K and 2.5M, respectively, when we go from 1 core to 24 cores on Edison. This step slows down by a factor of 2 when $\dnnz(\mathbf{x})$ is 200 because the overhead of managing $96$ buckets (24 threads multiplied by 4) becomes more expensive than performing the per-bucket merging operations.
Consequently, bucketing step starts to dominate the runtime of the SpMSpV-bucket algorithm on high concurrency.
The scalability of all components improves as the input vector becomes denser, as expected.

\section{Conclusions and Future Work}

We presented a work-efficient parallel algorithm for the sparse matrix-sparse vector multiplication (SpMSpV) problem. We carefully characterized 
different potential ways to organize the SpMSpV computation and identified
the requirements for a high-performance algorithm that is work-efficient and one that also avoids unnecessary synchronization.

Our algorithm avoids synchronization by performing a row-wise partitioning of the input matrix on the fly, and attains work efficiency by employing
the common computational pattern of the column-wise algorithms. 
Our algorithm achieves high-performance for a wide range of vector sparsity levels thanks to its vector-driven nature. 
The implementation of our algorithm on the Intel Ivy Bridge and the Intel KNL processors significantly outperforms existing approaches when the input vector
is very sparse, and performs competitively when the input vector gets denser
Matrix-driven algorithms are only competitive when the input vector gets relatively dense. As future work, we will investigate when and if it is beneficial to 
switch to a matrix-driven algorithm. 

Further refinements of the SpMSpV problem arise in different contexts. Some SVM implementations shrink the working set periodically, hence requiring
a data structure that is more friendly for row deletions. This could effect the tradeoffs involved in choosing the right SpMSpV algorithm, depending on the 
frequency of the shrinking. In addition, GraphBLAS effort is in the process of defining masked operations, including SpMSpV. This could also effect the 
algorithmic tradeoffs involved.  Studying those effects are subject to future work.

\section*{Acknowledgments}
This work is supported by the Applied Mathematics Program of the DOE Office of Advanced Scientific
Computing Research under contract number DE-AC02-05\-CH\-11231.
We used resources of the NERSC supported by the Office of Science of the DOE
under Contract No. DE-AC02-05CH11231.

\bibliographystyle{IEEEtran}
\bibliography{SpMSpV}

\end{document}